\documentclass[aps,prl,twocolumn,superscriptaddress,showpacs,draft]{revtex4}
\usepackage{graphicx}
\input{epsf}

\newcommand{\rem}[1]{}

\begin{document}

\title{Quantum synchronization and entanglement of two qubits \\ 
coupled to  a driven dissipative resonator}
\author{O.V.Zhirov}
\affiliation{\mbox{Budker Institute of Nuclear Physics,
630090 Novosibirsk, Russia}}
\author{D.L.Shepelyansky}
\affiliation{\mbox{Universit\'e de Toulouse, UPS, 
Laboratoire de Physique Th\'eorique (IRSAMC), F-31062 Toulouse, France}}
\affiliation{\mbox{CNRS, LPT (IRSAMC), F-31062 Toulouse, France}}

\date{April  2, 2009}


\begin{abstract} 
Using method of quantum trajectories we study the behavior of two 
identical or different superconducting qubits 
coupled to a quantum dissipative
driven resonator. Above a critical coupling
strength the qubit rotations become synchronized
with the driving field  phase and their evolution
becomes entangled even if two qubits may
significantly differ  from one another.
Such entangled qubits can radiate entangled photons
that opens new opportunities for entangled wireless communication
in a microwave range.
\end{abstract}
\maketitle

Recently, an impressive experimental progress has been
reached in realization of a strong coupling regime of 
superconducting qubits with a microwave resonator
\cite{wallraff,majer,wallraff1,wallraff2,wallraff3}. 
The spectroscopy of exchange of one to few microwave photons
with one \cite{wallraff,wallraff1}, two \cite{majer,wallraff2} 
and even three \cite{wallraff3}
superconducting qubits has been demonstrated to be
in agreement with the theoretical predictions
of the  Jaynes-Cummings and Tavis-Cummings models \cite{jaynes,tavis,scully}
even if certain nonlinear corrections have been visible.
Thus the ideas of atomic physics and quantum optics
find their promising implementations with 
superconducting macroscopic circuits
leading also to achievement of single artificial-atom lasing
in a microwave range \cite{astafiev}.

In comparison to quantum optics models \cite{jaynes,tavis,scully}
a new interesting element of superconducting qubits is
the dissipative nature
of coupled resonator which opens new perspectives for
quantum measurements \cite{korotkov}.
At the same time in the strongly coupled regime 
the driven dissipative resonator
becomes effectively nonlinear due to interaction
with a qubit that leads to a number of 
interesting properties \cite{shnirman1,gambetta,zhirov2008}.
Among them is  synchronization of qubit with
a driven resonator phase \cite{zhirov2008}
corresponding to a single artificial-atom lasing
realized experimentally in \cite{astafiev}.
The synchronization phenomenon has  broad
manifestations and applications in physics, engineering,
social life and other sciences \cite{pikovsky}
but in the above case we have a striking example
of quantum synchronization of a purely quantum qubit
with a driven resonator having a semiclassical
number of a few tens of photons   \cite{zhirov2008}.
This interesting phenomenon appears above a
certain critical coupling threshold
\cite{zhirov2008} and its investigation 
is  now in progress \cite{shnirman2}. 

Due to the experimental progress with two and three qubits
\cite{majer,wallraff2,wallraff3}
it is especially interesting to study  
the case of two qubits where an interplay of
quantum synchronization and entanglement 
opens a new field of interesting questions.
The properties of entanglement for two atoms (qubits)
coupled to photons in a resonator 
has been studied recently 
in the frame of Tavis-Cummings model within 
the rotating wave approximation (RWA) \cite{deutsch,solano}.
Compared to them we are interested in the case
of dissipative driven resonator strongly
coupled to qubits where RWA is not necessarily valid
and where the effects of quantum synchronization 
between qubits and the resonator is of primary importance.
In addition we consider also the case of
different qubits which is 
rather unusual for atoms but is very natural for 
superconducting qubits.

In absence of dissipation the whole system 
is described by the Hamiltonian
\begin{eqnarray}
\label{eq1}
&&   \hat{H}= \hbar\omega_0 (\hat{n}+1/2) + \hbar \Omega_1\sigma^{(1)}_z/2 +
   \hbar \Omega_2\sigma^{(2)}_z/2 \\
\nonumber 
&& +g \hbar\omega_0 (\sigma^{(1)}_x + \sigma^{(2)}_x) (\hat{a}+\hat{a}^+) 
   +f \cos \omega t \cdot(\hat{a}+\hat{a}^+)
\end{eqnarray} 
where the first three terms represent photons in a resonator 
and two qubits, $g$-term gives the coupling between qubits
and photons and the last term is the driving of resonator.
In presence of dissipation the resonator  dissipation rate
is $\lambda$  and its quality factor is assumed to be
$Q=\omega_0/\lambda \sim 100$. The decay rate of qubits
is supposed to be zero corresponding to a reachable 
experimental situation 
\cite{wallraff,majer,wallraff1,wallraff2,wallraff3}
where their decay rate is much smaller than $\lambda$.
The driving force amplitude
is expressed as $f=\hbar \lambda \sqrt{n_p}$
where $n_p$ is a number of photons in a resonator
at the resonance $\omega=\omega_0$ when $g=0$.
The whole dissipative system is described by the master equation for the
density matrix $\hat{\rho}$
which has the standard form \cite{weiss}:
\begin{equation}
\dot{\hat{\rho}} = -  \frac{i}{\hbar} [\hat{H},\hat{\rho}] +
\lambda ( \hat{a}  \hat{\rho} \hat{a}^{\dag} 
-  \hat{a}^{\dag}  \hat{a} \hat{\rho}/2
- \hat{\rho} \hat{a}^{\dag}  \hat{a}/2) .
\label{eq2} 
\end{equation}
The numerical simulations are done by the method of 
quantum trajectories \cite{percival} with
the numerical parameters and techniques described in \cite{zhirov2008}.

\begin{figure}
\centerline{\epsfxsize=8.5cm\epsffile{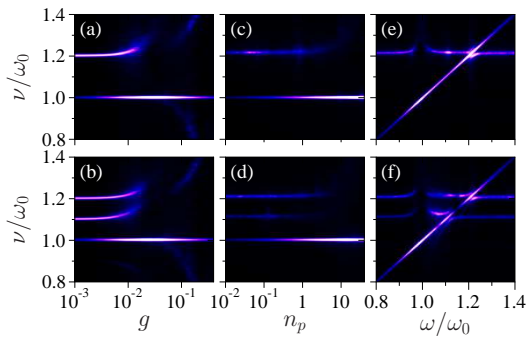}}
\vglue -0.2cm
\caption{(Color online) Spectral density $S(\nu)$ of two
driven qubits as a function of system
parameters for identical $\Omega_1/\omega_0=\Omega_2/\omega_0=1.2$
(top row) and different $\Omega_1/\omega_0=1.1$, $\Omega_2/\omega_0=1.2$ 
(bottom row) qubits: (a,b) $n_p=15$, $\omega/\omega_0=1$;
(c,d) $g=0.04$, $\omega/\omega_0=1$; (e,f) $n_p=15$, $g=0.04$.
Here and below $\lambda/\omega_0=0.02$.
}
\label{fig1}
\end{figure}

To analyze the system properties we
determine the spectral density of driven qubits defined as
$ S(\nu) = \left| \int dt \exp\{-i\nu t\} 
{\rm Tr}(\hat\rho (\sigma^{(1)}_x+\sigma^{(2)}_x)/2) \right|^2$.
Its dependence on system parameters for identical and 
different qubits
is shown in Fig.~\ref{fig1}. At small couplings $g$
the spectrum of qubits $S(\nu)$ 
shows the lines at the  internal qubit
frequencies $\Omega_{1,2}$ but above a certain critical
coupling strength $g>g_c$ the quantum synchronization
of qubits with the driven resonator takes place and
the unperturbed spectral lines are replaced
by one dominant spectral line at the driving frequency 
with $\nu=\omega$ (Fig.~\ref{fig1}a,b). 
A similar phenomenon takes place for $g>g_c(f)$ when
the strength of  resonator driving $f \propto \sqrt{n_p}$
is increased (Fig.~\ref{fig1}c,d). 
Indeed, with the growth of $n_p$
the number of photons in the resonator increases 
that leads to a stronger coupling 
between photons and qubits and eventual synchronization.
The synchronization of qubits with the resonator
is also clearly seen from  Fig.~\ref{fig1}e,f, 
where the spectral line of $S(\nu)$ follows firmly 
the variation of driving frequency $\omega$.
The striking feature of Fig.~\ref{fig1} is that even
rather different qubits with significant 
frequency detunings 
$|\Omega_{1,2}-\omega_0| \gg \lambda$
become synchronized due 
to their coupling with the resonator
getting the same lasing frequency $\nu \approx \omega$.

\begin{figure}
\centerline{\epsfxsize=8.5cm\epsffile{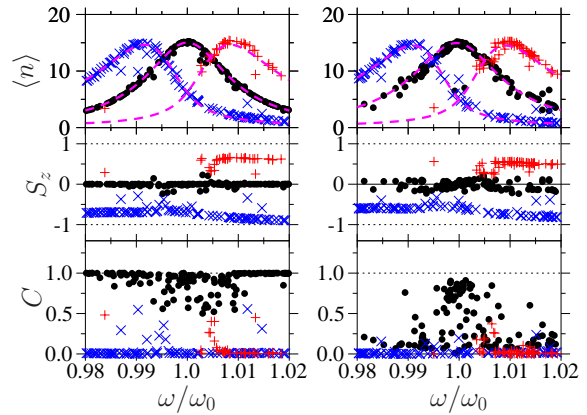}}
\vglue -0.2cm
\caption{(Color online) The average quantities of number of photons 
in the resonator $< n >$ (top), total spin
polarization $S_{z}$ (middle),  and 
the concurrence of two qubits $C$ (bottom) vs.
the driving frequency $\omega$. Left: identical qubits 
with $\Omega_{1}/\omega_0=\Omega_2/\omega_0=1.2$,
right: different qubits with 
$\Omega_1/\omega_0=1.1$, $\Omega_2/\omega_0=1.2$.
Other parameters are 
$\lambda/\omega_0=0.02$,  $n_p=15$, $g=0.04$. 
Symbols mark the values of the total spin
$S_z$:  $S_z > 0.2$ (red/gray $+$),
$|S_z| \leq 0.2$ (black dots),
$S_z < -0.2$ (blue/black $\times$).
Dashed curves on top panels show the resonance dependence
$n_{S_z}(\omega)$ (see text) with the resonance shift
$\Delta \omega_{S_z} = 
1.35 S_z g  \omega_0 \sqrt{(1-<S_z>^2)/(n_{S_z}+1)}$.
}
\label{fig2}
\end{figure}

For a better understanding of this phenomenon
we analyze the dependence of the average number 
of photons in the resonator  $< n >$ on
driving frequency $\omega$ 
(see Fig.~\ref{fig2}, top panels).
It has three pronounced maxima
which up to quantum fluctuations 
correspond to three values of 
the total spin component 
$S_z =  {\rm Tr}(\hat\rho (\sigma^{(1)}_z +
\sigma^{(2)}_z)/2)  $ being 
close to the values $S_z = -1, 0, 1$
with the total spin $S=1$ (triplet state)
as it is clearly seen from the data
shown in the middle panels of Fig.~\ref{fig2}. 
The whole dependence of $<n>$ on $\omega$
is well fitted by the resonance curves
$n_{S_z}=n_p \lambda^2/[4(\omega-\omega_0 -
\Delta \omega_{S_z})^2+\lambda^2]$  
where the frequency shift
$\Delta \omega_{S_z}$ 
appears due to the effective Rabi frequency $\Omega_R$
which gives
oscillations between the resonant states.
The value of $\Omega_R$ is induced by the 
coupling between 
qubits and photons in Eq.(\ref{eq1}) \cite{jaynes,tavis,scully}
and
can be approximated obtained as
an average value of coupling  that gives
$\Omega_R = 2 a S_z g \omega_0 \sqrt{(1-<S_z>^2)(n_{S_z}+1)}
 \approx 2 S_z a \Omega_{R0}$ with
$\Omega_{R0} = g \omega_0 \sqrt{n_{S_z}+1}$.
This gives  the frequency shift
$\Delta \omega_{S_z}=d \Omega_R/d n_{S_z}$
which determines the resonant dependence $ n_{S_z}(\omega)$
in a self consistent way. Such a theory gives a good description
of numerical data as it is shown in 
Fig.~\ref{fig2} (top panel)
where the corresponding values of $<S_z>$ are taken from the
middle panel. The numerical coefficient $a$ smoothly varies between
$1.14$ and $1.57$ for $0.01 \leq g \leq 0.06$. 
This resonance dependence is similar to the one qubit case
discussed in \cite{zhirov2008} but the 
effects of quantum fluctuations 
are larger due to mutual effective coupling
between qubits via the dissipative resonator.
On the basis of these estimates it is
natural to assume that the quantum
synchronization of a qubit
with a driving phase takes place under the condition
that the detuning is smaller than the 
typical value of Rabi frequency
\begin{equation}
\left| \Omega_{1,2}-\omega \right|
< \Omega_{R0} = g \omega_0 \sqrt{n_{S_z}+1} \approx
g \omega_0 \sqrt{n_{p}+1} .
\label{eq3}
\end{equation}
This criterion assumes a
semiclassical nature of photon field with
the number of photons $n_p > 1$.
Its structure is similar to the classical
expression for the synchronization 
tongue which is proportional to the 
driving amplitude being 
independent of dissipation rate \cite{pikovsky}.
The relation (\ref{eq3}) determines the border
for quantum synchronization 
$g_c \approx \left| \Omega_{1,2}-\omega \right|/\omega_0 \sqrt{n_{p}+1}$.
\begin{figure}
\centerline{\epsfxsize=8.5cm\epsffile{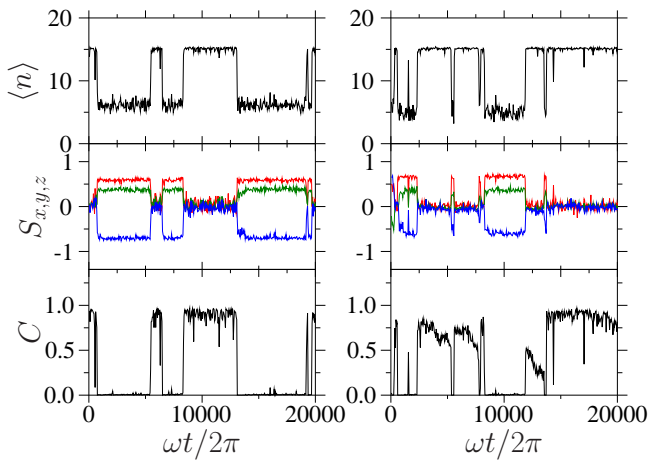}}
\vglue -0.2cm
\caption{(Color online)
Time evolution of system quantities: average photon number $<n>$ 
(top); components of total spin polarization $S_{x,y,z}$ (middle) 
where $x,y,z$-components  
correspond to red, green,
blue  colors (curves from top to bottom
at $\omega t/2\pi=1.5 \times 10^4$ (left) and 
$\omega t/2\pi=10^4$ (right));  concurrence $C$ (bottom). 
Left: two identical qubits with $\Omega_{1,2}/\omega_0=1.2$, right: 
two different qubits 
with  $\Omega_{1}/\omega_0=1.1$ and $\Omega_{1}/\omega_0=1.2$.  
Data are shown at stroboscopic moments of time 
with driving phase $\varphi = \omega t \; ({\rm mod} \; 2\pi) =0$,
Here $\omega/\omega_0=1$, other parameters  are as in Fig.~\ref{fig2}.
}
\label{fig3}
\end{figure}

The entanglement between qubits is 
characterized by concurrence $C$
(see e.g. the definition in \cite{deutsch}).
Its dependence on $\omega$
is shown in the bottom panels of Fig.~\ref{fig2}. 
It is striking that the concurrence $C$ 
can be close to unity not only for identical qubits
but also for different qubits.
Qualitatively, this happens due to synchronization
of qubits induced by the resonator driving
which makes them ``quasi-identical''
and allows to create the entangled state with $S_z=0$.

\begin{figure}
\centerline{\epsfxsize=8.5cm\epsffile{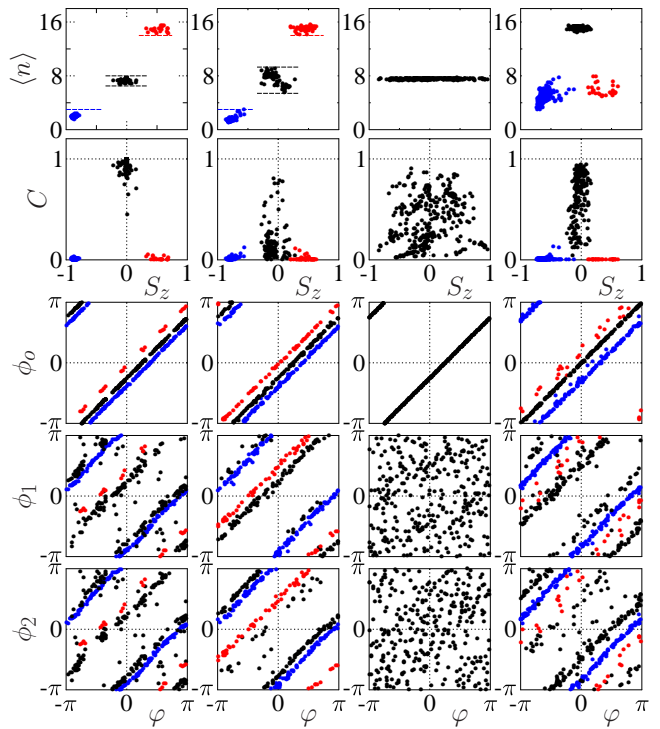}}
\vglue -0.2cm
\caption{(Color online) Rows from top to bottom:
     average photon number $\left\langle n \right\rangle $ {\it vs.} $S_z$;
     concurrence $C$ {\it vs.} $S_z$;
     resonator phase $\phi_o$, and qubits phases 
     $\phi_{1,2}$ {\it vs.} the driving 
     force phase $\varphi$. Columns from left to right: 
     (i) two identical qubits with 
     $\Omega_{1,2}/\omega_0=1.2$,  $g=0.04$, 
     (ii-iii) two different qubits, with $\Omega_1/\omega_0=1.1$, 
     $\Omega_2/\omega_0=1.2$, $g=0.04$ and  $g=0.001$, 
     all tree columns correspond to driving
     frequency $\omega/\omega_0=1.01$. The last column (iv) 
     shows data for two different qubits with
     $\Omega_1/\omega_0=1.1$, $\Omega_2/\omega_0=1.2$, $g=0.04$ 
     but the driving frequency $\omega/\omega_0=1$ ({\it cf.} (ii)).
     Colors of points are explained in the text. 
     Other parameters of simulations are:
     $\lambda/\omega_0=0.02$,  $n_p=15$.
}
\label{fig4}
\end{figure}

To understand the properties of the system in a better way we show
the time evolution of its characteristics along a typical quantum trajectory
in Fig.~\ref{fig3}. Average number of photons in the resonator
$<n>$ shows tunneling transitions between two metastable
states induced by quantum fluctuations (top panels). 
There are no transitions to the third metastable state,
seeing in Fig.~\ref{fig2} with three resonant curves,
but we had them on longer times or for other realizations
of quantum trajectories (see also Fig.~\ref{fig4}).
The life time inside each metastable state 
is of the order of thousands of driving periods
and the change of $<n>$ is macroscopically large 
(about a ten of photons). 
The transition leads also to a change of 
total spin polarization components
$S_{x,y,z} =  {\rm Tr}(\hat\rho (\sigma^{(1)}_{x,y,z} +
\sigma^{(2)}_{x,y,z})/2)$ (middle panels).
It takes place on a relatively 
short time scale $ \sim 1/\lambda$. 
The transition also generates emergence or death of concurrence
$C$ (or entanglement) 
which  happens on the same time scale $1/\lambda$
(bottom panels).
Naturally, $C$ is maximal when $S_x \approx 0$.
Remarkably, during long time intervals $C$ remains 
to be close to its maximal value $C=1$ even for
the case of different qubits. We attribute this phenomenon
to synchronization of two qubits by driven resonator.

To display and characterize this phenomenon in more detail
we determine the phases of oscillator and qubits via relations
$\phi_0=\arctan({\rm Im} \langle \hat{a} \rangle /
{\rm Re} \langle \hat{a} \rangle   )$,
$\phi_{1,2}=\arctan(<s^{(1,2)}_x>/<s^{(1,2)}_y>)$
respectively. The variation of these phases with the
driving phase $\varphi=\omega t \; ({\rm mod} \; 2\pi)$
is shown in Fig.~\ref{fig4} for different values of system
parameters corresponding to columns (i, ii, iii, iv).
For identical qubits (column (i))
the numerical data obtained along one long 
quantum trajectory form three well defined
groups of points in the plane of $<n>$ and $S_z$
(data are taken at stroboscopic 
moments of time $t$ with a certain frequency 
comparable but incommensurate with $\omega$
to sweep all phases $\varphi$).
It is convenient to mark these groups by three different
colors corresponding to small (blue or dark grey),
medium (black) and large (red or light grey)
values of $<n>$ (the groups are also marked by
horizontal lines). Such a classification shows that 
three groups have not only  distinct values of $<n>$
but also three  distinct locations in   concurrence $C$
and spin $S_z$.
Also these groups show three lines in the phase plane
for oscillator $(\phi_0,\varphi)$
and for each qubit $( \phi_{1,2},\varphi)$.
Of course, due to quantum fluctuations
there are  certain fluctuations for
qubit phases but the linear dependence between 
phases is seen very clearly,
thus showing the quantum synchronization
of system phases with the driving phase $\varphi$.
Physically, the three 
groups correspond to the three triplet states of total spin $S=1$.
Indeed, for identical qubits 
the states with total spin values  $S=1$ and $S=0$
are decoupled and the dynamics of $S=0$
state is trivial (see Eq.~(\ref{eq1})). 
For different qubits (columns (ii), (iv)) the quantum synchronization
between phases is also clearly seen even 
if two qubits have rather different frequency detunings.
For the case (ii) the concurrence is smaller compared to the
case of identical qubits (i) but by a  change of
driving frequency $\omega$ it can be increased (see the case (iv))
to the values as high as for identical qubits in (i).
We note that for different qubits (e.g. for (ii))
there is a visible splitting of the middle group of
black points in $(<n>,S_z)$ plane which
corresponds to states  with mixed components of
total spin $S=1$ and $S=0$:
indeed, for different qubits the coupling between
the states $S=1$ and $S=0$ is nonzero and such transitions
can take place (however, on the phase planes
the splitting of black points is too weak to be seen
in presence of quantum fluctuations).
Of course, the numerical data show the presence of
quantum fluctuations around straight lines in the phase planes.
Nevertheless, this regime of quantum synchronization
at $g> g_c$ is qualitatively different from the regime below
the synchronization border $g< g_c$ where
the points are completely scattered over the whole 
phase plane (column (iii)). In this regime (iii)
the qubits rotate independently from the 
resonator which stays at fixed number of photons.
In contrast, for $g>g_c$ two qubits move in 
quantum synchrony during a large number of
oscillations being entangled. 
A single superconducting qubit lasing has been
already achieved in experiments \cite{astafiev}.
Our theoretical studies show that 
such two qubits, which in practice are always
non-identical, can be made entangled and 
produce lasing in synchrony with each other.
Being entangled such qubits can radiate
entangled photons in a microwave range.
Therefore, the experiments similar to \cite{astafiev}
but with two single-atoms lasing would
be of great interest for entangled microwave
photons generation. 

In conclusion, our numerical simulations show that
even two different superconducting qubits 
can move in quantum synchrony induced by coupling
to a driven dissipative resonator, which can make them
entangled. Such entangled qubits can radiate 
entangled microwave photons 
that opens interesting opportunities for 
wireless entangled communication in a
microwave domain.

The work is funded by  EC  project EuroSQIP
and RAS grant  "Fundamental problems of nonlinear dynamics".
\vglue -0.2cm

\end{document}